\documentclass[aps,prl,twocolumn]{revtex4}
\def \be {\begin{equation}}
\def \ee {\end{equation}}
\def \bea {\begin{eqnarray}}
\def \eea {\end{eqnarray}}

\newcommand{\eq}[1]{(\ref{#1})}
\def\a{\alpha}      
\def\b{\beta}       
    
\def\d{\delta}

\def\m{\mu} \def\n{\nu}
\def\o{\omega}
 
\def\r{\rho}

\def \a {\alpha}
\def \b {\beta}

\def \d {\delta}
\def \eps {\epsilon}
\def \m {\mu}
\def \n {\nu}

\def \r {\rho}
\def \o {\omega}

\begin{document}
\title{Transplanckian Entanglement Entropy}
\author{Darwin Chang$^{1}$, Chong-Sun Chu$^{1,2}$ and Feng-Li Lin$^{3}$}
\affiliation{$^{1}$ Department of Physics, National Tsing-Hua University,
Hsin-Chu, 300 Taiwan}
\affiliation{$^{2}$ Department of Mathematics, University of Durham, DH1 3LE,
United Kingdom}
\affiliation{$^{3}$ Physics Division,
National Center for Theoretical Sciences,
National Tsing-Hua University, Hsin-Chu, 300 Taiwan}

\begin{abstract}
The entanglement entropy of the event horizon is known to be
plagued by the UV divergence due to the infinitely blue-shifted
near horizon modes. In this paper we calculate the entanglement
entropy using the transplanckian dispersion relation, which has
been proposed to model the quantum gravity effects. We show that,
very generally, the entropy is  rendered UV finite due to the
suppression of high energy modes effected by the transplanckian
dispersion relation.
\end{abstract}
\maketitle
\newpage

After 30 years of study, the microscopic origin of the
Bekenstein-Hawking entropy \cite{BH} of a Schwarzschild black hole
remains mysterious.
This thermodynamic entropy is finite and is proportional to the
area of the event horizon.
Matter field contribution to the entropy can be calculated by
tracing over the fluctuations beyond the horizon, and is known as
the quantum entanglement entropy. The entanglement entropy for
black hole was first calculated by 't Hooft \cite{hooft} by using
the so called brick wall method, later on, it was generalized to
arbitrary bounded hypersurface \cite{Sred} and to Rindler space
\cite{Susskind,CW1}. In \cite{solo} a covariant method was
developed.
In all these results, the entanglement entropy
obeys an  area law. However, they all suffer from
the usual UV divergence of quantum field theory in curved spacetime.
This UV divergence is  related \cite{Susskind} to the Hawking information
paradox.

Intuitively the UV divergence can be understood to be caused by
the infinitely blue-shifted near horizon modes. One of the
underlying assumptions in these calculations is that the
energy-momentum relation takes the same form up to arbitrary high
energies. However there are convincing reasons to believe that the
conventional picture of spacetime structure, as well as spacetime
symmetries, will get modified or even lose its meaning at energy
scale comparable to the Planck scale. In a viable theory of
quantum gravity, e.g. string theory, there is no UV divergence. It
is therefore natural to expect that quantum gravity effects (maybe
together with other effects) will suppress the transplanckian
modes and hence improve the UV behavior of the entanglement
entropy, or even cure the divergences.
Although there is so far no proof of it from a theory of quantum
gravity,  it has been proposed in \cite{Unruh,Corley} that one may
model the quantum gravity effect on matter propagation by
modifying the dispersion relation such that the transplanckian
modes are suppressed. In this letter we show that there is a
generic class of dispersion relations with such suppression which
indeed gives rise UV finite entanglement entropy!

We first review the WKB brick wall method introduced by 't Hooft
\cite{hooft} and apply it to calculate the entanglement entropies
for Rindler space, black hole (BH) and the de Sitter space. Then
we show that within a general class of transplanckian dispersion
relation(TRD), the entanglement entropy is UV finite.  The
condition that this class of TRD has to satisfy is shown to be
actually very generic.  We then give explicit results for the
entanglement entropy based on the TRD proposed by Unruh
\cite{Unruh} as example.  Finally we conclude with a few remarks
about the implications of our results.

{\it Entanglement Entropy and UV divergence.---} Consider a
massless scalar in a 4-dimensional  curved spacetime described by
the metric $g_{\m\n}$ with an event horizon
located at $r_h$. The usual dispersion relation is
$ g_{\m\n}p^{\m}p^{\n}=0\;. $
The entanglement entropy was found to be divergent
due to the infinitely blue-shifted
near horizon region \cite{Sred,hooft}. In the brick wall model, a
UV cutoff is introduced
to regularize the divergence
by imposing the condition on the wavefunction
$ \Phi(r)=0$ for $r \le r_h+\epsilon$,
where $\epsilon$ is
the infinitesimal brick wall thickness \cite{hooft}. This has the
effect of cutting out the near horizon modes.

The total free energy is obtained by summing over the
contributions from all the physical modes satisfying the brick wall
condition,
\be \label{free1}
\b F
= \int d\omega \; z(\beta \omega) {dg(\omega) \over d\omega}
\ee
where the Boltzmann weight $z(x):=\ln(1-e^{-x})$, $\beta$ is
the inverse temperature,
and $g(\omega)$ is the density of states. In the WKB
approximation \cite{hooft,Susskind},  it is
\be
\label{g} \pi g(\omega) =\int dr \int { A \;d^2 p_{\perp}\over
(2\pi )^2}\;p_r (\omega,\vec{p}_{\perp},r),
\ee
for the Rindler case. Here $A$ is the
(regulated) area of the horizon,
$r$ and $p_r$ is the radial coordinate and its conjugate
momentum, and $p_{\perp}$ is the transverse momentum.
As we will see shortly, the integration limits are determined by
the dispersion relation and the
above-mentioned brick wall condition.
Finally, the entanglement entropy can be
obtained from the free energy as
$ S=\beta^2 {\partial F}/{ \partial \beta}$.

To be generic, consider a spherically symmetric background metric
of the form \be \label{metric} ds^2 = g_{00} dt^2 + g_{rr} dr^2 +
f d\Omega_2^2, \ee where $g_{00}, g_{rr}$ and $f$ are functions of
$r$. The dispersion relation is \be \label{disperse}
g^{00}\omega^2+g^{rr}p^2_r+p_{\perp}^2=0\;, \ee where $g^{00}
=1/g_{00}, g^{rr}= 1/g_{rr}$ and $p_{\perp}^2=g^{mn}p_mp_n =
\ell(\ell +1)/f$ is the transverse momentum squared
and $\ell$ is the angular momentum quantum number. It is
convenient to introduce the following definitions \be
\label{new-def} \xi^2 := p_\perp^2 g_{rr}= {\ell(\ell
+1)g_{rr}\over f}, \quad \rho^2 := - {g_{00} \over g_{rr}}, \ee
the dispersion relation \eq{disperse} takes the form \be p_r
=\sqrt{{\omega^2 / \rho^2}-\xi^2}. \ee
The density of state, $\pi g(\o) =\int  dr d\ell\; (2\ell+1) \;
p_r$, then takes the form \be\label{density} \pi g(\o) = \int
d\rho \mu(\rho) d\xi \;  \xi \; p_r, \ee where $\mu(\rho)$ is a
``measure factor'' defined by \be \label{meas} \mu(\rho) := {2 f
\over g_{rr}}{dr\over d\rho} . \ee
For fixed $\omega$ and $\rho$, we require $\xi \le \omega/\rho$ to
have $p_{r}$ real. This condition should be imposed in the
$\xi$-integration.

For pedagogical purpose, we now consider some specific cases.
First, the Rindler space described by the metric \be ds^2=-r^2
dt^2+ dr^2 + dx^2+ dy^2\;. \ee The event horizon of the local
patch is located at $r=0$. In this case, $\rho=r$, $\xi^2=
p_\perp^2$, $f=A/4\pi$,
and the measure factor $\mu(\rho)=A/(2\pi)$.
%
Therefore
\be
\label{rindlerd} g(\omega)={A\over 2\pi^2}
\int^{\infty}_{\epsilon} d\rho \int^{\omega/\rho}_0 d\xi\; \xi \;
p_r = {A \omega^3 \over 12 \pi^2 \epsilon^2} .
\ee
Here a brick wall thickness is introduced in the lower limit of
$\rho$-integration.
Together with Eq.(\ref{free1}), we reproduce the result in
\cite{Susskind}, i.e.,
\be \label{rdfs}
F=-{A\pi^2 \over 180 \beta^4 \epsilon^2}, \quad
S= \frac{A}{360 \pi \epsilon^2},
\ee
which are plagued by the UV
divergence as $\epsilon \rightarrow 0$. In the evaluation of $S$,
we have set $\beta =2 \pi$ for the inverse Rindler temperature.
In \cite{Susskind} the UV divergence was interpreted as an
renormalization of the Newton constant.

Using the same method we can also calculate
the free energy of the black hole
background
\be
\label{bh1} ds^2=-h(r)dt^2+h(r)^{-1}dr^2+r^2 d\Omega_2^2,
\ee
where $0< h(r)=1-{ 2M /r} < 1$.
In this case $\rho=h$ and
$\mu(\rho)=2(2M)^3 \rho/(1-\rho)^4$.
The density of state is
\bea
g(\omega)
&=& {2 (2M)^3 \over \pi} \int^{1-\delta}_{\epsilon} d\rho {\rho
\over (1-\rho)^4} \int^{\omega /\rho}_0 d\xi\; \xi\; p_r \;\;\;\;
\\
&=&{2 (2M)^3\omega^3 \over 3 \pi} ({1\over \epsilon} + {1\over 3
\delta^3}+\cdots) ,  \label{g-bh} \eea where we have introduced an
UV (resp. IR) cutoff $\epsilon$ (resp. $\delta$) in the
$\rho$-coordinate, which corresponds to an UV (resp. IR) cutoff
$r=2M/(1-\epsilon)$ (resp. $r= 2M/\delta$ ) in the $r$-coordinate;
and $\cdots$ are the sub-leading terms in the limit of $\delta,
\epsilon \rightarrow 0$. Using (\ref{g-bh}) and $\beta = 8 \pi M$,
we get
\be \label{hooft-bh} F={-2\pi^3 (2M)^3 \over 45 \beta^4}({1
\over \epsilon} + {1\over 3\delta^3}), \quad
S={1\over 360}({1 \over \epsilon} + {1\over 3\delta^3})
\ee
which is plagued by both UV and IR divergences.
This agrees with the eqn. (3.12) of \cite{hooft} except that a IR
divergent boundary contribution had been dropped there. The
IR-divergent piece is independent of $M$ and represents the
contribution from the vacuum. In \cite{hooft}, the UV-divergent
piece was shown to give an entropy which obeys the area law
$S = {A\over 360\pi \epsilon^2_{\rm p}}$
if the UV-cutoff is
given in terms of the proper distance, i.e.
$\epsilon_{\rm p}\approx 4M\sqrt{\epsilon}$.
This is the same form as \eq{rdfs}.
In modern language, the relation between $\eps$ and $M$
for fixed $\epsilon_{\rm p}$
required by the area law indicates a holographic nature of UV/IR
connection \cite{UVIR}.

Finally we consider the example of de Sitter space
described by the metric \eq{bh1} but with
$0< h(r)=1-r^2/R^2 \le 1$. In this case,
we have again $\rho=h$, but
the measure factor is now $\mu(\rho)=R^3 \rho\sqrt{1-\rho}$.
We find the same result \eq{rdfs}
if the UV-cutoff is in terms of $\epsilon_{\rm p} \approx
R\sqrt{\epsilon}$.


In conclusion: The entanglement entropy associated with the event
horizon obeys the universal area law, $S = A/ (360 \pi \epsilon^2_{\rm p})$, for
the UV cutoff $\eps_{\rm p}$, and diverges as $\eps_{\rm p}
\rightarrow 0$.

{\it Transplanckian entropy.---} The above UV divergence is due to
the infinite contribution of the near horizon modes which have
energy far beyond the Planck scale. It is quite generally believed
that nonlocal effects due to quantum gravity will provide a
natural regulator to the UV divergence in quantum field theory by
suppressing the contributions of the high energy modes. It is
therefore very natural to expect that in a consistent theory of
quantum gravity, the  above results will become finite. The
simplest proposal to encode the effects of this suppression is to
replace the linear dispersion relation by the so-called
transplanckian dispersion relation(TDR) \cite{Unruh,Corley}.

Consider a spherically symmetric background
given by the general metric \eq{metric},
it is reasonable to impose the TDR according to the residual
spacetime symmetry preserved by the metric.  Since the
transplanckian effect will be mostly due to the blue-shift in the
near horizon regime in the radial direction, so it is natural to
impose TDR along that direction to suppress the blue-shift effect.
Therefore we consider TDR of the form
\be \label{disperse2}
g^{00}\omega^2+g^{rr}H^2(p_r)+p_{\perp}^2=0\;.
\ee
The function $H$ describes the transplanckian effects.
In terms of the $\rho$ and $\xi$ variables, the TDR \eq{disperse2}
then takes the form
\be \label{p-H}
p_r= H^{-1} (\sqrt{{\omega^2/ \rho^2}-\xi^2} )\;.
\ee

For example,  Unruh \cite{Unruh} and respectively, Corley and
Jacobson \cite{Corley} (C-J) proposed
\bea \mbox{Unruh}:&
H^U_n(k)=k_0[\tanh({k / k_0})^n]^{1\over n}\;,
\label{disp-unruh} \\
\mbox{C-J}: & H^{CJ}(k)=\sqrt{k^2-{k^4/( 4k_0^2)}}, \quad k \le 2
k_0\;. \label{disp-corley}
\eea
The corresponding radial momentum is
\bea
\mbox{Unruh}:& p_r =k_0 \{\tanh^{-1}[k_0^{-2}({\omega^2/ \rho^2}
-\xi^2)]^{n\over
2}\}^{1\over n}, \; \label{p-unruh} \\
\mbox{C-J}: & p^{\pm}_r=\sqrt{2} k_0\sqrt{1\pm \sqrt{1-k_0^{-2}
({\omega^2/ \rho^2}-\xi^2)}}. \;\;\; \label{u-corley}
\eea
Both Unruh's and C-J's proposals belong to a general class of TDR
in which the suppression of high energy modes is effected by a
bounded function $H$, says,
\be\label{bound}
0\le H^2(k) \le k_0^2,
\ee
for some $k_0$.
It follows that for this bounded class of TDR
\be \label{energy} \max({\omega^2/ \rho^2}-k_0^2,0) \le \xi^2 \le
{\omega^2 /\rho^2}\;.
\ee
Combining with the brick wall condition
$\rho\ge \epsilon$, we have quite generally the following
integration branches contributing to the free energy
(up to a factor of $1/(\b \pi)$)
\bea
&(1)&
\int_{\epsilon k_0}^{\omega_{\rm m}} d\omega  \;z
\frac{d}{d\omega} \int_\epsilon^{\omega / k_0} d\rho \m(\rho)
\int_{\sqrt{{\omega^2/\rho^2}-k_0^2}}^{\omega / \rho} d\xi \xi p_r,
\;\;  \label{br1} \\
&(1*)& \int_{(1-\delta) k_0}^{\infty} d\omega \; z
\frac{d}{d\omega} \int_\epsilon^{1-\delta} d h  \m(h)
\int_{\sqrt{{\omega^2 / h^2}-k_0^2}}^{\omega / h} d\xi \xi p_r ,
\;\;\;\;\;\; \label{br1s}\\
&(2)&  \int_{\epsilon k_0}^{\omega_{\rm m}} d\omega \;z
\frac{d}{d\omega} \int_{\omega / k_0}^{\rho_{\rm m}}  d\rho\m(\rho)
\int_0^{\omega/\rho} d\xi\xi p_r , \label{br2}\\
&(3)& \int_0^{\epsilon k_0} d\omega \;z  \frac{d}{d\omega}
\int_{\epsilon}^{\rho_{\rm m}}  d\rho \m(\rho)
\int_0^{\omega/\rho} d\xi \xi p_r,
\label{br3}
\eea
where $\rho_{\rm m}=\omega_{\rm m}=\infty$ for Rindler case,
$\rho_{\rm m}=1-\delta$, and $\omega_{\rm m}=(1-\delta)k_0$ for
BH and de Sitter case.
Note that by comparing the ranges of the above $\xi$-integrations
with the ones without transplanckian suppression, we find that the
use of TDR leads to a reduction of the allowed angular momentum
modes. This leads to a suppression on the density of states and
hence the UV finiteness of the entropy as shown below.

It is convenient to change variable $x=\sqrt{\o^2/\rho^2-\xi^2}$
in order to isolate the $\o$-dependence in $g(\o)$.
We have
\be
\label{br}
\pi g(\o)=  \int_{\cdot}^{\cdot} d\rho\;
\mu(\rho)\int_{\cdot}^{\cdot} dx \;x H^{-1}(x)\;.
\ee
Note that the integrand does not depend on
$\o$ and $\eps$,
but the integration limits may.  Note also that the
$(1*)$-branch only exists for BH and de Sitter due to the
additional constraint $h\le 1$.

{\it UV Finiteness.---} Our goal is to determine the UV
behavior of $F$ and $S$ for a generic bounded TDR with a $H$
satisfying \eq{bound} such that there are the branches
\eq{br1}-\eq{br3}. One may try to estimate the UV behavior for the
integrals for each branches. However this is rather complicated
technically. Surprisingly, without knowing explicitly these
integrals, one can show that the sum of the branches (1), (1*),
(2) and (3), i.e. the free energy $F$, is completely UV finite!

To demonstrate this, let us first consider (1*) branch. The
integration limits are independent of $\o$, i.e.
$\int^{1-\delta}_{\eps} d\rho \int^{k_0}_0 dx$, so is $g(\o)$.
Therefore (1*) does not contribute. For (1) branch the
$dx$-integral is independent of $\o$, however, there is $\o$ dependence
for $g(\o)$ coming from the integration limits for
$d \rho$-integral. Using the fundamental theorem of calculus, we can carry
out the derivative w.r.t $\o$ in \eq{br1} and obtain
\be \label{F1}
F_{(1)}={1\over \pi \b k_0}\int^{k_0}_0 dx x H^{-1}(x) \cdot
\int^{\o_{\rm m}}_{\epsilon k_0} d\o \;z(\b\o)\;\mu({\o\over k_0}).
\ee
Similarly, we can carry out the derivative w.r.t. $\o$ for
branch (2) and (3), and obtain
\be
\label{F2}
F_{(2)}=-F_{(1)}+{1 \over \pi \b}\int^{\o_{\rm m}}_{\epsilon k_0}
d\o z \int^{k_0}_{\o \over \rho_{\rm m}} dx\; \mu({\o\over x})
H^{-1}(x)\;,
\ee
\bea
\label{F3}
F_{(3)}&=&{1\over \pi \b}\int_0^{\epsilon k_0} d\o z
\int_{k_0}^{\o\over \eps}dx\; \mu({\o\over x})H^{-1}(x)\nonumber\\
&+&{1\over \pi \b}\int_0^{\epsilon k_0} d\o z
\int^{k_0}_{\o\over \rho_{\rm m}} dx\; \mu({\o\over x})H^{-1}(x)\;.
\eea
Physically it is required that the integral \be \label{H-cond}
I:=\int^{k_0}_0 dx\; x H^{-1}(x) < \infty \ee to be finite. The
reason is because the density of states for branch (1), in
\eq{br1}, is given by  $\pi g(\o) = I \cdot \int_\epsilon^{\omega
/ k_0} d\rho \m(\rho) $, and it should be finite for any sensible
physical system. This condition also guarantees our manipulation
above for \eq{F1}, \eq{F2} to make sense. Of course \eq{H-cond} is
true for both  Unruh's and C-J's TDRs.

Next we examine the first line of \eq{F3}, we perform a change of variables
by $\omega=\epsilon k_0 y$ and obtain
\be \label{small}
{k_0 \eps \over \pi
\beta}\int^1_0 dy \; z(\beta\epsilon k_0 y) \int^{k_0 y}_{k_0} dx \;
\mu({\epsilon k_0 y\over x})
H^{-1}(x) \;.
\ee
Using the fact $z(x)\approx \ln(x)-x/2+\cdots$ around $x=0$,
and assuming the following condition on the near horizon geometry
\be\label{meas-cond}
\eps \mu(\eps)\approx \epsilon^\a \quad\mbox{for}\quad \a>0,
\quad \mbox{as}\quad \eps \to 0,
\ee
which is true for all three metrics considered, we can
manipulate \eq{small} by interchanging the order of integrations
over $dx$ and $dy$ so that the $y$-integration can be carried out,
\eq{small} then becomes
\be
{\eps^\a \ln \eps \over \pi \a k_0} \int^{k_0}_0 dx\; x H^{-1}(x)\;.
\ee
This vanishes as $\eps \to 0$.

Finally, the second line of \eq{F3} can be combined with $F_{(2)}$
of \eq{F2}, and we obtain for the total free energy
\be \label{final}
F={1\over \pi \b}\int^{\o_{\rm m}}_0 d\o\; z(\b\o)
\int^{k_0}_{\o\over \rho_{\rm m}} dx\; \mu({\o\over x})H^{-1}(x).
\ee
Note that $F$, and hence $S$, is independent of $\eps$, i.e. UV
finite. This result is quite general and is true as long as
{\bf (i)} the transplanckian modification of the radial modes
takes the form \eq{p-H} with the suppression condition \eq{bound},
{\bf (ii)} the metric satisfies a near horizon condition, which
expressed in terms of the measure factor as \eq{meas-cond}.
We thus see that TDR generally yields a UV finite free energy
and entanglement entropy. This is the main result of this paper.

{\it Examples.---}  Now we would like to present some more
explicit results. We will consider the Unruh dispersion relation
with $n=1$ as an example. For the Rindler case, all the integrals
can be worked out explicitly for all three branches. The results
are
\bea \label{ext1} F_{(1)}&&= {Ak_0^2\over 4 \pi^2
\beta^2}\int^{\infty}_{\beta \epsilon k_0}
dy \ln(1-e^{-y}) \approx - \frac{A k_0^2}{24 \beta^2}, \\
F_{(2)} &&= (2 \ln 2 -1) F_{(1)},
\eea
where the approximation in (\ref{ext1}) is justified by the
infinitesimal $\beta \eps k_0$, and
\be
F_{(3)} =\frac{A k_0^3 \epsilon}{4 \pi^2 \beta} \int_0^1 dx
\ln(1-e^{-{\beta k_0 \epsilon x}}) [x \ln\frac{1+x}{1-x} +
\ln(1-x^2)] .\nonumber
\ee
It is easy to show that, with
$k_0, \beta$ fixed, $F_{(3)}\to 0$, as $\epsilon \to 0$. Therefore
the total free energy and entropy are \be \label{freeRd} F= -
\frac{\ln 2}{12}\frac{ A k_0^2 }{\beta^2 } ,\qquad S={\ln2 \over
12 \pi} Ak_0^2. \ee Note that $k_0$ plays the role of an UV
regulator.
In fact with
$k_0\to c/ (\beta \epsilon)$, the total free energy is of the form
$F=(A/\eps^2)\beta^{-4} f(c)$. For $c=3.4645$, $f(c)=-\pi^2/180$,
the result reduces to (\ref{rdfs}).
This is expected since $H^U_1(k)\stackrel{k_0 \rightarrow
\infty}{\longrightarrow} k$.
Moreover, as long as $k_0$ is far
less than the Planck energy scale, the entanglement entropy is
just a small correction to the Bekenstein-Hawking entropy.

For the case of BH, the integrals cannot be evaluated for each
individual branches. Nevertheless our general result \eq{final}
gives the leading IR piece
\be\label{xx}
F \approx  -{2k_0^3  \over 3 \pi \beta} ({2M\over \delta})^3
\int^1_0 dy \;y \ln(1-e^{-\beta k_0 y}) \tanh^{-1} y
\ee
where we assume a large IR cutoff $2M/\delta(\gg \beta)$ to
extract the leading IR term from the $h$-integration. The IR
divergence is independent of the background $M$ as in the
non-transplanckian case. We note that
eqn. \eq{xx} is of the form $F= (2M/\d)^3\b^{-4} g(\b k_0)$ with
$\lim_{\a\to\infty} g(\a) = -2\pi^3/135$. Therefore it reproduces
the result \eq{hooft-bh} either for $k_0 \to \infty$ with $\b$
fixed or for the zero temperature limit $\b\to\infty$ with $k_0$
fixed. In general, by performing a numerical study on \eq{xx}, we
find that, for fixed $k_0$, both the free energy and the entropy
are monotonically decreasing functions of $\beta$.

For the case of de Sitter, eqn. (\ref{final}) gives a result
of the form $F=(R^3/\beta^4)j(\beta k_0)$ which
is free of both UV and IR divergences. However, $j(x)$ is a
complicated function and yields no area law for the entanglement
entropy even in the low temperature limit, i.e. $\beta=2\pi R
\rightarrow \infty$.
The results for the Corley-Jacobson
dispersion relation is similar. We will skip it here.

{\it Discussions.---} In this letter we show that the free energy
and the entanglement entropy for a wide class of spacetime (so
long as the finiteness condition on the ``measure factor''  is
satisfied) are UV finite when the high energy modes are suppressed
with the use of TDR.  What we show is that
for a certain class of TDR, the density of states at high energies
does get reduced to a sufficient degree such that the entanglement
entropy is UV finite.
This is quite a remarkable result because a priori it is
not obvious at all.
We would like to emphasis that this UV finiteness is largely
due to a modification of the propagator, which can be think of as an
{\it UV regulator},  high momentum modes do not need to be {\it cut out}.

Our results also show that, with the quantum correction, the entanglement
entropy may have a complicated
area dependence and does not conform to a simple area law in
general.  This is expected since quantum gravity corrections
generally induces higher derivative gravity correction terms that
modify the area law, see for examples \cite{deWit,Carlip}.
It is natural that such modification is also seen
in the matter part; and we have shown that this is indeed the
case.

We note that without transplanckian suppression, the near horizon modes
reproduce the UV behavior of the full integration.  However, this is
not the case in the transplanckian case. If one keeps only the
$\rho=\epsilon$ region in the $\r$-integral in evaluating the free
energy, we find that the near horizon modes do yield UV improved
results, which is one order less
divergent than the ordinary case.
However it does not render the result finite, at least for the
Rindler case, as is given by the full integration.
This indicates that the nonlinear suppression feature of the
TDR plays a crucial role in improving the UV behavior.

As we mentioned, the TDR of
interests are typically characterized by a suppression scale $k_0$
as in \eq{bound}. We remark that the transplanckian scale $k_0$
plays the role of a UV regulator.
In fact we have shown for the Rindler case that,
by taking $k_0 \sim (\b \eps)^{-1}$, the ordinary UV divergent
result is recovered.
One can show this similarly for the generic case.
%

It deserves further study on the thermodynamical behavior of these
transplanckian QFT systems and on the origin of the
TDR in the context of string theory, see for example \cite{cgs}
for a discussion.
Also it is interesting to re-examine the information paradox
based on our result and the earlier study of non-thermal
transplanckian Hawking radiation \cite{Brout,Corley}.

\begin{acknowledgments}
We thank Bin Chen, Ko Furuta, Pei-Ming Ho, Ted Jacobson  and
Hoi-Lai Yu  for discussions and comments. We acknowledge grants
from Nuffield foundation of UK, NCTS and NSC of Taiwan.
\end{acknowledgments}

\end{document}